\documentclass{article}

     \usepackage[final]{neuripsDeepInverse_2021}

\usepackage[utf8]{inputenc} 
\usepackage[T1]{fontenc}    
\usepackage{hyperref}       
\usepackage{url}            
\usepackage{booktabs}       
\usepackage{amsfonts}       
\usepackage{nicefrac}       
\usepackage{microtype}      
\usepackage{amsmath}
\usepackage{xcolor}         
\usepackage[ruled, linesnumbered]{algorithm2e}
\usepackage{setspace}
\usepackage{wrapfig}
\usepackage{graphicx}
\usepackage{float}
\usepackage{graphbox}
\usepackage{multirow}
\usepackage{tikz}
\usepackage{stackengine}
\usepackage{caption}
\usepackage{subcaption}

\newcommand{\bbzero}{\text{\usefont{U}{bbold}{m}{n}0}}
\MakeRobust{\bbzero}

\usetikzlibrary{arrows}
\tikzstyle{block} = [rectangle, draw, fill=blue!20, outer sep=5pt,
text width=4em, text centered, rounded corners, minimum height=4em]

\title{A Closer Look at Reference Learning\\for Fourier Phase Retrieval}

\author{%
	Tobias Uelwer \\
	Department of Computer Science\\
	Heinrich Heine University\\
	Düsseldorf, Germany \\
	\texttt{tobias.uelwer@hhu.de} \\
	\And
	Nick Rucks \\
	Department of Computer Science\\
	Heinrich Heine University\\
	Düsseldorf, Germany \\
	\texttt{nick.rucks@hhu.de} \\
	\And
	Stefan Harmeling \\
	Department of Computer Science\\
	Heinrich Heine University\\
	Düsseldorf, Germany \\
	\texttt{stefan.harmeling@hhu.de} \\
}

\begin{document}

\maketitle

\begin{abstract}
Reconstructing images from their Fourier magnitude measurements is a
problem that often arises in different research areas. This
process is also referred to as phase retrieval. In this work, we
consider a modified version of the phase retrieval problem, which
allows for a reference image to be added onto the image before the
Fourier magnitudes are measured. We analyze an unrolled
Gerchberg-Saxton (GS) algorithm that can be used to learn a good
reference image from a dataset. Furthermore, we take a closer look
at the learned reference images and propose a simple and efficient
heuristic to construct reference images that, in some cases, yields
reconstructions of comparable quality as approaches that learn
references. Our code is available at \url{https://github.com/tuelwer/reference-learning}.
\end{abstract}

\section{Introduction}

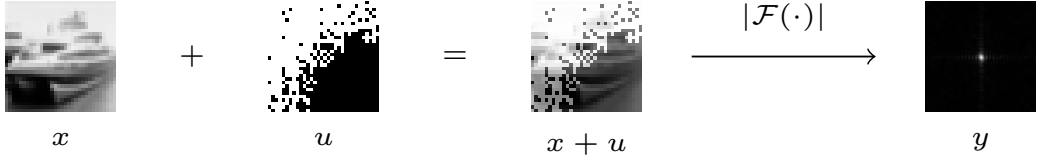
\begin{figure*}
	\centering
	\resizebox{1\textwidth}{!}{
		\begin{tikzpicture}
			\node[inner sep=1pt, ,
			label=below:{\tiny$x$}](image) at (0, 0) {\includegraphics[width=.06\textwidth]{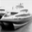}};
			\node[inner sep=1pt, label=below:{\tiny$u$}]
			(ref) at (2 , 0) {\includegraphics[width=.06\textwidth]{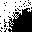}};
			\node[inner sep=1pt, label=below:{\tiny$x+u$}]
			(ref-and-image) at (4, 0) {\includegraphics[width=.06\textwidth]{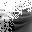}};
			\node[inner sep=1pt, label=below:{\tiny$y$}]
			(magn) at (7 , 0) {\includegraphics[width=.06\textwidth]{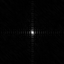}};
			\draw[->] (4.8, 0) -- (6.2, 0);
			\node at (5.5,0.3) {\tiny{$|\mathcal{F}(\cdot)|$}};
			\node at (1,0) {\tiny{$+$}};
			\node at (3,0) {\tiny{$=$}};
	\end{tikzpicture}}
	\caption{An overview of the measurement process with a known reference.}
	\label{fig:measurement}
\end{figure*}

In general, Fourier phase retrieval is the problem of reconstructing an image from its Fourier magnitude measurements. The problem arises in different research areas, e.g., in X-ray crystallography \cite{millane1990phase}, astronomical imaging \cite{fienup1987phase}, optics \cite{walther1963question}, array imaging \cite{bunk2007diffractive}, or microscopy \cite{zheng2013wide}. 
In particular, for non-oversampled measurements, phase retrieval has not been fully solved yet. In this work, a slightly modified version of the problem is considered: instead of reconstructing the image from the plain magnitude measurements, a reference image is added onto the original image before the Fourier magnitudes are measured, i.e., we reconstruct the image $x\in \mathbb{R}^{d\times d}$ from the modified measurements
\begin{equation}
	y = |\mathcal{F}(x+u)|,
\end{equation} where $\mathcal{F}$ denotes the discrete two-dimensional Fourier transform and $u\in \mathbb{R}^{d\times d}$ is a known reference image. Furthermore, we assume that both the image and the reference are non-negative. Figure \ref{fig:measurement} gives an overview of the measurement process, that can also be implemented in practice.

\section{Related Work and our Contributions}

The modified phase retrieval problem was first analyzed by Kim and Hayes~\cite{kim1990phase, kim1990iterative}. Recently, Hyder et al.~\cite{hyder2020solving} showed that such a reference can be
learned from a rather small dataset using an unrolled gradient descent algorithm.  While it is interesting that this is possible, a couple of questions arise which we discuss in this work.
\paragraph{Our Contributions:}
\begin{enumerate}
	\item We modify the well-established Gerchberg-Saxton (GS) algorithm~\cite{gerchberg1972practical} so that it utilizes a reference.  
	Furthermore, we show that this modified GS algorithm can be unrolled to learn a reference image (similar to the approach of Hyder et al.~\cite{hyder2020solving}).
	\item We answer the question under which conditions learning a reference image for Fourier phase retrieval is really necessary. For this, we propose a simple baseline reference image that is easily constructed and compare its performance with the learned references in the oversampled and the non-oversampled case.
\end{enumerate}
The idea of unrolling algorithms has been considered before for other phase retrieval problems: The performance of an unrolled GS algorithm for phase retrieval has been analyzed by Schlieder et al.~\cite{schlieder2020learned}.  Naimipour et al.~\cite{naimipour2020unfolded} unrolled the Wirtinger flow algorithm for compressive phase retrieval.  Unlike our work, these papers augment existing phase retrieval algorithms with learnable parameters which is different from the reference based phase retrieval we consider in this work.

\section{Unrolling the Gerchberg-Saxton Algorithm to Learn a Reference}
\label{sec:unrolling}

The phase retrieval algorithm discussed in Hyder et al.~\cite{hyder2020solving} is a gradient descent method that requires extensive parameter tuning of its step size.  For this reason, we unroll the GS algorithm, which typically converges within a few iterations.  The GS algorithm iteratively replaces the magnitude of the current estimate with the measured magnitude and enforces a positivity-constraint on the image after the reference has been subtracted. Algorithm~\ref{alg:gerchberg-saxton} shows our modified version of GS that utilizes a reference image. 
We learn the reference by calculating the mean squared error (MSE) between outputs of the GS algorithm after $n$ iterations and the corresponding original images.  Then we perform backpropagation to calculate the gradient with respect to the reference $u$ and update it using a gradient descent step. These steps are repeated for multiple batches until convergence is achieved.

\IncMargin{1em}
\DontPrintSemicolon
\SetAlFnt{\normalsize}
\SetAlCapFnt{\normalsize}
\SetAlCapNameFnt{\normalsize}
\begin{algorithm}[H]
	\setstretch{1.4}
	\KwIn{Fourier magnitude $y\in \mathbb{R}^{d\times d}$, reference image $u\in \mathbb{R}^{d\times d}$, initialization $x_{0}\in\mathbb{R}^{d\times d}$, number of iterations $n$}
	\KwOut{Reconstruction $x_n\in \mathbb{R}^{d\times d}$}
	\For{$k=1,\dots,n-1$}{
		$p_{k+1}\gets \mathcal{F}(x_{k}+u)/|\mathcal{F}(x_{k}+u)|$ \tcp*{Estimate phase}
		
		$\bar x_{k+1}  \gets \mathcal{F}^{-1}(p_{k+1} \odot y)-u$ \tcp*{Fourier constraints and subtract u}
		
		$x_{k+1}\gets\max(0, \bar x_{k+1})$ \tcp*{Image constraints (max is element-wise)}
	}
	\textbf{return} $x_n$ 
	\DecMargin{1em}
	\caption{Gerchberg-Saxton algorithm for phase retrieval with a reference image}
	\label{alg:gerchberg-saxton}
\end{algorithm}
\DecMargin{1em}

\section{Constructing Simple References Mimicking the Learned Ones}
\begin{figure}
	\centering
	\begin{tabular}{ccc} 
		\scriptsize Learned ref. from Hyder et al.~\cite{hyder2020solving} 
		& \scriptsize Learned ref. for GS (ours) 
		& \scriptsize Simple ref. (ours)\\
		\includegraphics[height=1.13cm, align=c]{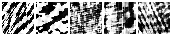}& \includegraphics[height=1.13cm, align=c]{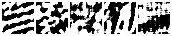} & \includegraphics[height=1.13cm, align=c]{figures/our_reference.png} 
	\end{tabular}
	\caption{Comparison of different learned references. For each of the two learned methods, the shown references are trained on the following datasest: MNIST, EMNIST, FMNIST, SVHN, CIFAR-10. Our simple reference is shown on the right.}
	\label{fig:references}
\end{figure}

Learning the reference image is computationally expensive.  
The most obvious baseline reference is a random image. For this, we sample the pixel values from a uniform distribution $\mathcal{U}(0,1)$. Looking at various learned reference images produced by the algorithm of Hyder et al.~\cite{hyder2020solving} and our unrolled GS algorithm (Figure~\ref{fig:references}), we see that most pixels are either zero or one.  This suggests to also consider random binarized reference images.

However, looking more closely at those learned reference images (Figure~\ref{fig:references}), we see that all references exhibit flat areas. Furthermore, the learned references do not show any symmetries. These observations suggest trying a simple reference image as a baseline mimicking those features.  To construct such a reference, we start with a black image that has a white square in the bottom right corner. After blurring this image with a Gaussian filter, we normalize and add weighted Poisson noise to every pixel. Finally, we binarize the image appropriately. A resulting simple reference is shown on the right in Figure~\ref{fig:references}.

In the next section, we compare the performance of these different references for reconstructing images from their oversampled and non-oversampled magnitude measurements.

\section{Comparing Baseline and Learned References}

\paragraph{Experimental Setup.}\begin{wrapfigure}[]{r}{0pt}
	\includegraphics[width=0.4\linewidth]{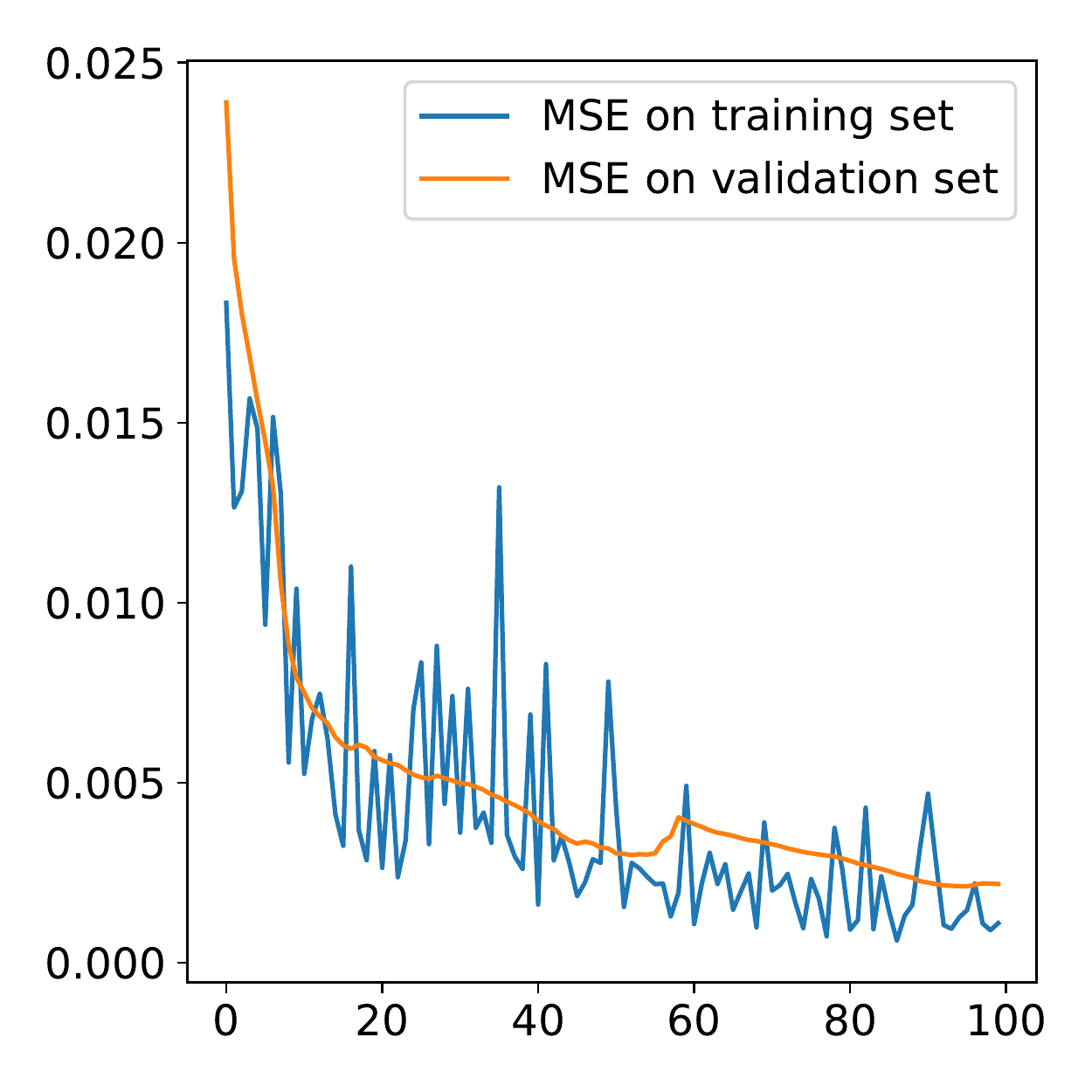}
	\caption{Training and validation MSE between reconstructions and original images of the FMNIST dataset.}
	\vspace{-10pt}
	\label{fig:training}
\end{wrapfigure}

To study whether references are necessary at all, we also reconstruct images from measurements that were taken without adding a reference onto the image. In that case, the trivial ambiguities (translation and flip) cannot be resolved by the reconstruction algorithm. Therefore, we register the reconstructions appropriately before calculating the error. For fair comparison, we register the reconstructions of the other methods before calculating the MSE as well.

We evaluate the reconstructions of images from the MNIST~\cite{lecun1998gradient}, the EMNIST~\cite{cohen2017emnist}, the FMNIST~\cite{xiao2017fashion}, the SVHN~\cite{netzer2011reading}, and the CIFAR-10~\cite{krizhevsky2009learning} datasets. We convert the images from the SVHN and the CIFAR-10 dataset to grayscale. 
We consider non-oversampled magnitude measurements as well as measurements that are oversampled by a factor of two in each dimension.
To solve the Fourier phase retrieval problem, we use a similar algorithm as Hyder et al.~\cite{hyder2020solving}.
We run the algorithm for $500$ steps and set the step size to $\alpha=1.95$, which was chosen on a validation dataset. 
For each of the datasets, we use $1000$ images for testing and we train the references on $100$ training images using  Adam \cite{kingma2014adam} with a learning rate of $0.01$. We use batches of size $10$. During training, we unroll the GS algorithm for $15$ steps and use $500$ steps to reconstruct the images at test time.

Similar to the intensities of the images, we restrict the entries of the reference image to lie between $0$ and $1$ as using larger values makes the problem trivial to solve, i.e., even a random reference with values between $0$ and $100$ yields perfect reconstructions.

\paragraph{Results.}

\begin{table}
	
	\begin{center}
		\begin{tabular}{l l c c c c c} 
			\toprule
			& Method & MNIST & EMNIST & FMNIST & SVHN & CIFAR-10\\
			\midrule
			\multirow{6}{*}{\rotatebox{90}{\scriptsize Non-oversampled}}& No reference & 0.035615 & 0.063414 & 0.042417 &  0.013985 & 0.035134\\
			& Random ref.&0.052724  & 0.079784 & 0.046670 &  0.012393 &0.030222\\
			& Random ref. (binary)   & 0.055324 & 0.081130 &  0.049436& 0.012447 & 0.029299\\ 
			& Simple ref. (ours)     & 0.060027 & 0.089347& 0.067549&0.011270 & 0.024128\\
			& Hyder et al.~\cite{hyder2020solving} & 0.002607 &\textbf{0.014687}&  \textbf{0.013649}& 0.010131 &0.024141 \\ 
			& Unrolled GS (ours)     & \textbf{0.002181}& 0.015427 & 0.019863&\textbf{0.008775} &\textbf{0.020020}\\ 
			\midrule
			\multirow{6}{*}{\rotatebox{90}{\scriptsize Oversampled}} & No reference & 0.020566&  0.032907&0.021068 & 0.007516 & 0.020518\\
			& Random ref.         & 0.005350 & 0.025308 & 0.011484 & 0.003670 & 0.009848\\
			& Random ref. (binary)& 0.001170 & 0.010994& 0.006842 & 0.002462 & 0.007310\\ 
			& Simple ref. (ours)  & 0.000761 & 0.003681 & 0.000848 & 0.000187 & 0.000495\\
			& Hyder et al.~\cite{hyder2020solving} ref. &0.000132  & \textbf{0.000023} & 0.000073 & 0.000126 & 0.001415\\ 
			& Unrolled GS ref. (ours) &\textbf{0.000071} & 0.000257 & \textbf{0.000055}& \textbf{0.000055}&\textbf{0.000125}\\
			\bottomrule
		\end{tabular}
	\end{center}
	\caption{Comparison of the mean squared reconstruction error using different references for Fourier phase retrieval from  non-oversampled and oversampled magnitudes. Best values are printed \textbf{bold}.}
	\label{tab:results}	
\end{table}

\begin{figure}
	\centering
	\setlength\tabcolsep{1pt}
	\begin{tabular}{ccccc} 
		& \scriptsize No reference& \scriptsize Simple ref. (ours) & \scriptsize Hyder et al.~\cite{hyder2020solving} ref. & \scriptsize Unrolled GS ref. (ours) \\
		\multirow{3}{*}{\rotatebox[origin=c]{90}{\scriptsize \hspace{-1.4cm}Non-oversampled}}&\includegraphics[height=1.13cm, align=c]{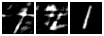} & \includegraphics[height=1.13cm, align=c]{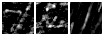}& \includegraphics[height=1.13cm, align=c]{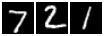} &
		\includegraphics[height=1.13cm, align=c]{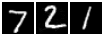} 	\\
		
		&\includegraphics[height=1.13cm, align=c]{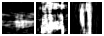}
		&\includegraphics[height=1.13cm, align=c]{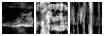}& \includegraphics[height=1.13cm, align=c]{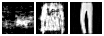} &
		\includegraphics[height=1.13cm, align=c]{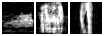} \\
		
		&\includegraphics[height=1.13cm, align=c]{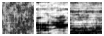}
		&\includegraphics[height=1.13cm, align=c]{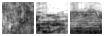}& \includegraphics[height=1.13cm, align=c]{figures/reconstructions/non-oversampled/reconstruction_cifar10_baseline.png} &
		\includegraphics[height=1.13cm, align=c]{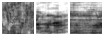}\\
		
		\midrule
		
		\multirow{3}{*}{\rotatebox[origin=c]{90}{\scriptsize \hspace{-1.4cm} Oversampled}}&		\includegraphics[height=1.13cm, align=c]{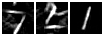} &	\includegraphics[height=1.13cm, align=c]{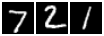}& \includegraphics[height=1.13cm, align=c]{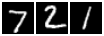} &
		\includegraphics[height=1.13cm, align=c]{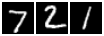} \\
		
		&\includegraphics[height=1.13cm, align=c]{figures/reconstructions/non-oversampled/reconstruction_fmnist_no.png} 
		&\includegraphics[height=1.13cm, align=c]{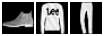}& \includegraphics[height=1.13cm, align=c]{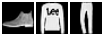} &
		\includegraphics[height=1.13cm, align=c]{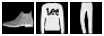} \\
		
		&\includegraphics[height=1.13cm, align=c]{figures/reconstructions/non-oversampled/reconstruction_cifar_no.png}
		&\includegraphics[height=1.13cm, align=c]{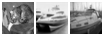}& \includegraphics[height=1.13cm, align=c]{figures/reconstructions/oversampled/reconstruction_cifar_baseline.png} &
		\includegraphics[height=1.13cm, align=c]{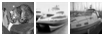}  \\
	\end{tabular}
	\caption{Reconstructions of the MNIST, the FMNIST and the CIFAR-10 dataset from  non-oversampled and oversampled Fourier magnitudes using different references.}
	\label{fig:reconstructions}
\end{figure}

Table~\ref{tab:results}	shows that our unrolled GS algorithm is able to learn references of similar quality as the method proposed by Hyder et al.~\cite{hyder2020solving}.
We see that the reconstruction errors of our simple reference are quite close to the errors of the learned methods in the oversampled case.
Figure~\ref{fig:training} shows the training and the validation loss for a reference trained using our unrolled GS algorithm on the FMNIST dataset. 
We can observe that only a few stochastic gradient descent steps are necessary to drastically decrease training and validation error.
Also, our GS algorithm requires only a small number of iterations in training to learn a good reference.
Figure~\ref{fig:reconstructions} visualizes some reconstructions.
In the non-oversampled case, the learned references produce better reconstructions than our simple reference.
Surprisingly, the random references that we use as baseline perform worse on the MNIST-like datasets than using no baseline in the non-oversampled case.
In the oversampled case, the learned references, as well as our simple reference, produce almost perfect reconstructions.  Therefore, we conclude that learning references is not necessary when the Fourier magnitudes are oversampled.

\section{Conclusion}
In this paper, we discuss how the GS algorithm can be unrolled in order to learn a reference image for phase retrieval. We show that our unrolled GS algorithm performs comparably to an existing method and requires only a few unrolling steps while having no hyperparameters that require extensive tuning.  Furthermore, we provide a simple reference which puts the benefit of a learned reference into perspective, especially when oversampled magnitude measurements are available.

\section{Broader Impact}
Fourier phase retrieval is a fundamental and relevant imaging problem in many areas of science, e.g., in X-ray crystallography or microscopy.  Ethical or societal consequences depend on the exact application of the phase retrieval algorithm.  
Our insights set the performance gain of learned reference images for phase retrieval into perspective. We propose a simpler and more efficient approach for reference construction which yields similar results and requires less computation and thus less energy.

\bibliographystyle{plain}
\bibliography{bibliography}

\end{document}